# P

## Potential of Augmented Reality for Intelligent Transportation Systems

Adnan Mahmood, Bernard Butler and
Brendan Jennings
Emerging Networks Laboratory,
Telecommunications Software & Systems Group,
Waterford Institute of Technology,
Waterford, Ireland

## Synonyms

Augmented reality; Intelligent transportation systems; Network resource management; Vehicular ad hoc networks; Virtual reality

## Definition

Augmented reality (AR) is usually regarded as a human-computer interface which enhances a human's perception of the real world by superimposing useful contextual information in real time (Ling 2017). Over the past few years, AR systems have made remarkable advances in several industries, including aerospace, automotive, robotics, health, military, education, marketing, social media and networking, tourism, gaming, etc. It is anticipated that the next revolutionary wave in industry would come in the shape of a significant dependence on AR devices and systems. In contrast, virtual reality (VR) is a computer-generated three-dimensional immersive experience which is displayed either on a computer display or via special stereoscopic displays, such as Facebook's *Oculus Rift* system.

## Introduction and Background

Rapid advances in wireless communication technologies coupled with ongoing massive development in vehicular networking standards and innovations in computing, sensing, and analytics have paved the way for intelligent transportation systems (ITS) to develop rapidly in the near future. ITS provides a complete solution for the efficient and intelligent management of real-time traffic, wherein sensory data is collected from within the vehicles (i.e., via their onboard units) as well as data exchanged between the vehicles, between the vehicles and their supporting roadside infrastructure/network, among the vehicles and vulnerable pedestrians – subsequently paving the way for the realization of the futuristic Internet of Vehicles (IoVs) (Yang et al. 2017). The traditional intent of an ITS system is to detect, monitor, control, and subsequently reduce traffic congestion based on a real-time analysis of the data pertinent to certain patterns of the road traffic, including traffic density at a geographical area of interest, precise velocity of vehicles, current and predicted travelling trajectories and times, etc.

However, merely relying on an ITS framework is not an optimal solution. In case of dense traffic environments, where communication broadcasts from hundreds of thousands of vehicles could potentially choke the entire network (and so could lead to fatal accidents in the case of autonomous vehicles that depend on reliable communications for their operational safety), a fall back to the traditional decentralized vehicular ad hoc network (VANET) approach becomes necessary. It is therefore of critical importance to enhance the situational awareness of vehicular drivers so as to enable them to make quick but well-founded manual decisions in such safety-critical situations. Modern-day vehicles that utilize sophisticated cameras and radars to project front/rear images and to detect obstacles, respectively, need to present this contextual information to drivers in ways that inform and do not distract them. Conventional displays mounted on dashboards or middle consoles divert the attention of drivers from the primary task of driving. Also, contextual information in these displays is not usually superimposed, and if superimposed, it is presented in a higher degree of detail which not only increases the cognitive load of the drivers but also adds to the overall delay of responding to an unforeseen event. This is where AR comes into play, which, unlike its predecessor (i.e., virtual reality), *enhances* a driver's perception by superimposing useful real-world context in minimal detail in a driver's line of sight.

AR is realized by executing the following four basic tasks (Glockner et al. 2014):

*Scene Capture*. In the first stage, reality is captured either by employing a video-capturing device (camera) or a see-through device (head-mounted display).

*Scene Identification*. Once the reality has been captured, it should be scanned in order to identify a particular position where the respective virtual content should be embedded. This position may be identified with virtual tags (markers) or tracking technologies such as sensors, laser, infrared, or GPS.

*Scene Processing*. Once the captured scene is completely identified, its corresponding virtual information needs to be requested from a database via the Internet. The associated information, with up to the minute detail, is accumulated and subsequently processed in the cloud (either at the network edge or remote cloud) with sufficient compute resources to filter the minimal useful virtual content.

*Scene Visualization*. Lastly, the AR system generates an amalgamated image of the real scene with the superimposed virtual content, which is subsequently displayed in the driver's visual field.

AR is regarded as one of the several key technology enablers which could set up the conditions for a "perfect technological storm" in the coming decades, as illustrated in Fig. 1. It is interesting to note that the video-processing aspects of AR, in convergence with some of these other technology enablers, could meet many of the more difficult requirements of future autonomous vehicles. The widespread adoption of such vehicles would eventually lead to the realization of next-generation ITS frameworks and smart cities. Apart from the automotive sector, AR is anticipated to revolutionize many industries, including manufacturing, healthcare, education, logistics, architecture, military, gamification, and data centers (Uchida et al. 2017).

AR requires specialized hardware and software, and the development and implementation of this software relies on a highly capable, state-of-the-art hardware and networking. These have been the subject of ongoing research and development by scientists and engineers in the domains of microprocessors, sensors, display systems, battery life, high-resolution cameras, mobile internet speeds, and more. At the time of writing (2018), the "industry best practices" in terms of available AR hardware include handheld devices (*smart phones and tablets*), stationary AR systems (*AR wardrobes*), spatial AR systems, head-mounted displays (*Oculus Rift*), smart glasses, smart contact lenses, etc.

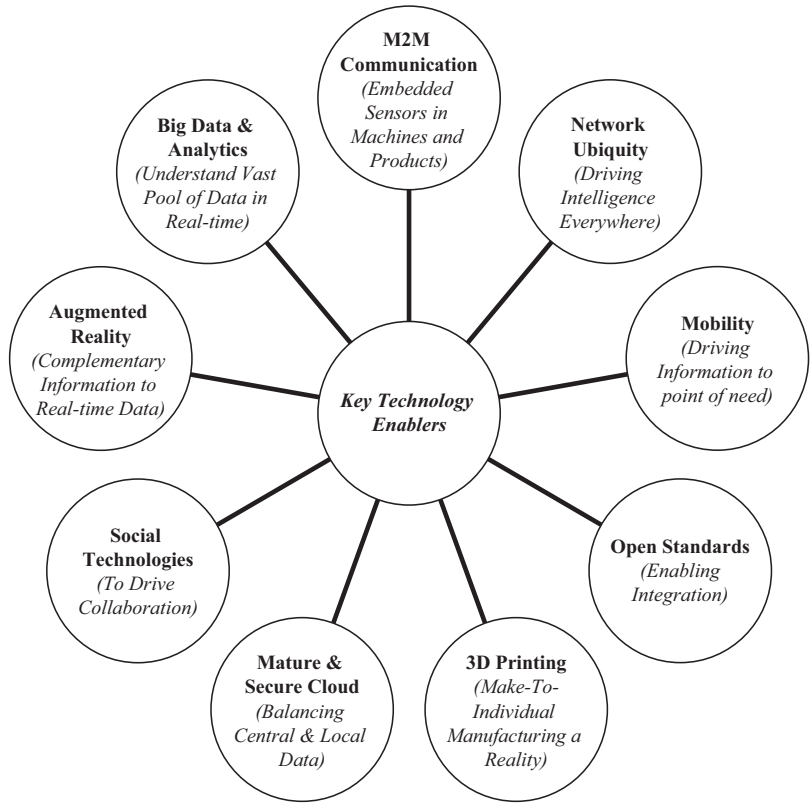

**Potential of Augmented Reality for Intelligent Transportation Systems, Fig. 1** Powering technological innovation – key technology enablers for Industry 4.0

## The Need and Potential of Augmented Reality for ITS

Since the advent of automobiles as early as the mid-nineteenth century, each vehicle has been driven by a single human driver, according to his/her requirements, preferences, etc. However, with massive increases in vehicular traffic leading to road congestion and soaring energy costs, the potential benefits of cooperative driving, especially "*platooning*" has received attention. Vehicle platooning aims to significantly reduce the inter-vehicular distances relative to that recommended for manual driving. This is achieved by (1) making driving more predictable by sharing common objectives rather than having individual drivers make uncoordinated decisions and (2) by partially or fully automating common driving tasks, offering faster reaction times than could be achieved by unassisted human drivers. Hence, the smaller the space between vehicles becomes, the more vehicles can be packed in that (finite) road space, thereby increasing road capacity. A supplementary benefit of closer vehicle placement is that it can lead to reduced aerodynamic drag, thereby increasing energy efficiency (Qiu et al. 2017).

Nevertheless, significantly reducing inter-vehicular distances also needs to be safe. This aspect (safety) has been studied in the ITS research literature and highlighted as one of the main challenges hindering the full realization of ITS. Accordingly, significant technological advances in autonomous vehicles and advanced driver assistance systems (ADAS) have been made over the years that have completely altered our perception of transportation safety. Google, Delphi, Tesla, Mercedes-Benz, Porsche, Nissan, and Bosch (to just name a few) are already testing safely driven autonomous vehicles and numerous intelligent ADAS functions, such as lane departure warning, collision warning, autonomous emergency braking, and adaptive cruise control. Some of these are now being offered as options in production vehicles (Kaur and Sobti 2017). Nevertheless, in spite of these technological developments, road accidents continue to happen due to human factors such as misjudgments and erratic human behavior. Research studies (Rameau et al. 2016; Abdi and Meddeb 2018) have analyzed the causes of fatal traffic accidents. Head-on collision as a result of overtaking at high speed is one of the major causes of road fatalities and generally occurs due to occlusion (blocking) of the overtaking driver's field of view, as depicted in Fig. 2. AR could be extremely beneficial here by enabling the overtaking driver to "see through" the slow-moving lorry to the vehicles immediately preceding it and thus make safer overtaking decisions.

Such enhanced situational awareness is highly valuable, given all the in-vehicle secondary tasks assigned to the driver, from adjusting the temperature and ventilation to conversing with other passengers. In the autonomous cars of the future, distractions will be even greater: infotainment for listening to music, making and/or receiving calls, reading and/or sending emails, etc. Also, the level of attention required while driving varies according to the conditions, as driving in dense fog or heavy rain and snow usually requires a higher degree of risk awareness compared to driving on a sunny day with a clear view. Thus, limited visibility and driver distractions pose grave threats to situational awareness and can cause serious road accidents. Also, providing a driver with excessive data results in higher cognitive workloads and weaker driving performance (Contreras et al. 2017). Furthermore, while current sophisticated driving systems can support drivers by providing detailed information, the presentation systems generally require drivers to divert their attention to displays mounted on dashboards or middle consoles. This is where AR comes into play by rendering *critically important* information in a driver's line of sight, thereby decreasing both cognitive workload and distraction while driving. However, it is pertinent to note that the information being displayed should be carefully chosen. It should be derived from a wide variety of onboard sensors, enhanced with data from the roadside and other vehicles, shared over wireless links, and intelligently extracted with sufficient but not excessive detail. Remote data can be shared using vehicle-to-vehicle (V2V), vehicle-to-infrastructure/network (V2I/N), and vehicle-to-pedestrian (V2P) communication. Also, it is essential to display this information with very low latency: the maximum tolerable delay in the case of safety-critical vehicular applications is in the range of 20–100 ms as depicted in Table 1.

## The Need for Intelligent Network Resource Management in ITS

The low-latency requirements of safety-critical vehicular applications increase the demand for intelligent and efficient network resource management, because vehicular AR applications have numerous parts that need to be managed. At present, the number of sensors on vehicles is approximately 100, and this number is expected to double by 2020. According to Intel estimates, autonomous vehicles are anticipated to generate and consume 5 TB of data for every hour of driving, and the primary reason for this appetite is the growth of onboard sensors with automotive cameras and radars alone generating 20–40 Mbps and 100 kbps of data, respectively (Nelson 2018). ITS networks also need to support additional data streams such as that generated by humans (vehicular users and roadside pedestrians). As estimated, vehicular users on average consume around 650 Mbps of videos, chat, and Internet usage per day, and this is anticipated to increase to 1.5 Gbps per day by 2020. The third source of data arises from certain crowd-sourced traffic platforms, such as Google Waze. Processing such big data arising from hundreds of thousands of vehicles requires extremely high computational power and intelligent machine learning algorithms to extract critically important information to be rendered for each driver.

Furthermore, unlike traditional networks, VANETs are highly distributed in nature, and owing to the high mobility of vehicles, the geospatial environmental context also changes rapidly, leading to a dynamically changing network topology. Traditional networking architectures were designed for more static contexts and are thus unable to cater for such challenges. Consequently, there has been a shift toward

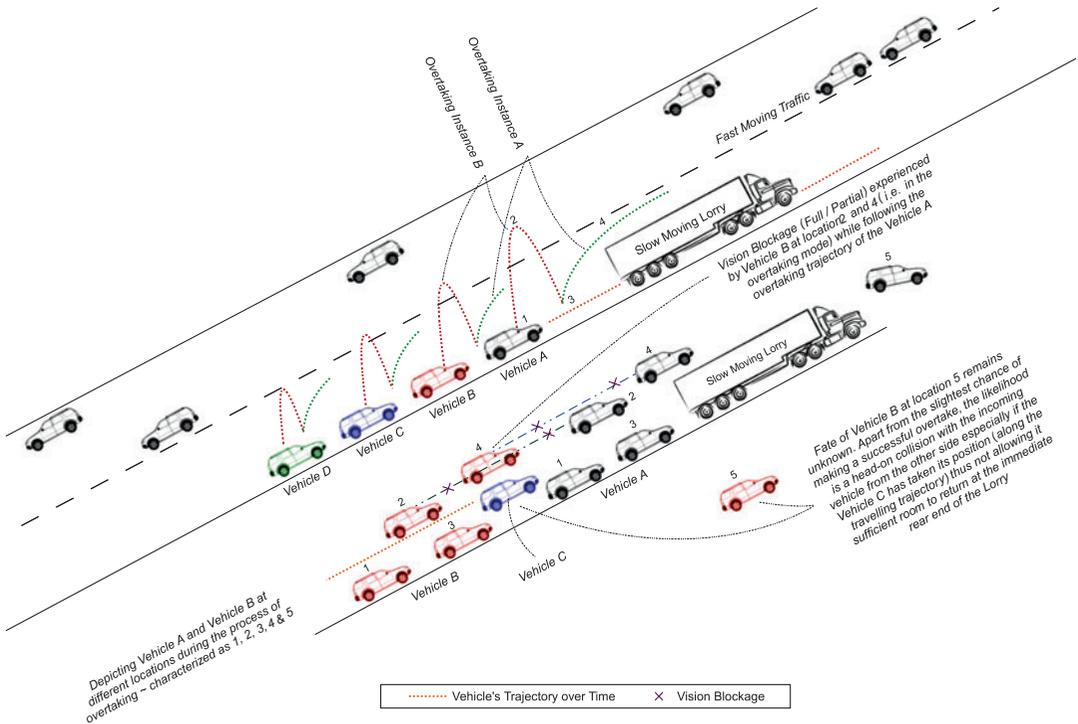

**Potential of Augmented Reality for Intelligent Transportation Systems, Fig. 2** Depicting human behavior in overtaking scenarios – conventional driving scenario

**Potential of Augmented Reality for Intelligent Transportation Systems, Table 1** Example link layer parameters for V2X services

| Scenarios | Effective range (meters) | Absolute speed of UE (km/h) | Relative speed b/w two UEs (km/h) | Tolerable latency (ms) |
|---|---|---|---|---|
| Major roads/suburban | 200 | 50 | 100 | 100 |
| Motorway/freeway | 320 | 160 | 280 | 100 |
| Autobahn | 320 | 280 | 280 | 100 |
| Non-line of sight/urban | 150 | 50 | 100 | 100 |
| Urban intersections | 50 | 50 | 100 | 100 |
| Shopping area/campus | 50 | 30 | 30 | 100 |
| Imminent crashes | 20 | 80 | 160 | 20 |

Source: 3GPP Technical Report 22.885 v14.0.0 (3GPP 2015)
It should be noted that there are various parameters involved in these scenarios, i.e., physical (relative speed, effective range), incidental (line-of-sight occlusion), and service-related (maximum tolerable latency). These parameters can be derived from the principle that there should be more than 95% probability that a vehicle has received at least one V2X message from two consecutive V2X messages notifying it of any impending incident.

a more agile approach to ensure reprogrammability, scalability, flexibility, and elasticity. The emerging and promising paradigm of software-defined networking (SDN) together with edge-based computing can help to meet these challenges. SDN decouples the data plane from the control plane, and network intelligence is passed on to a logically centralized controller that is responsible for automating and orchestrating the network operations. Edge-based computing brings network resources, i.e., both compute and storage, nearer to the edge and proves extremely useful when caching frequently requested safety-critical and non-safety (i.e., infotainment) applications and services, thereby reducing network delay by mitigating excessive network management overhead.

## Conclusion

In most countries, the use of private motor vehicles has been on the rise, as consumers are drawn to the flexibility and convenience of such vehicles. Given associated trends such as increasing urbanization with the potential for increasing road congestion, the need for safer urban road environments has become more acute. Thus, the management of urban traffic demands a balance between throughput, safety, efficiency, and sustainability. Private motor vehicles need to share the roads with other modes of transport (i.e., public transport, bicycles, and pedestrians); this also adds to the complexity. Every road accident is one too many, especially when it results in severe injuries or the loss of human life. According to one estimate (Oh et al. 2016), more than 80% of road accidents worldwide happen due to the negligence of drivers or other human errors. Though there have been many sophisticated road safety technologies introduced over the past few years to enhance road safety, many of them resulted in distraction or cognitive overload for the drivers concerned. There is only a limited amount of visual stimuli to which drivers can attend, understand, and react without taking their eyes off the road, and it is essential to present drivers with only critical contextual information in minimal detail within their line of sight so that the right decisions can be made at the right time. This manuscript outlines the significance and need of AR in general and especially for the ITS domain and indicated some of the networking requirements necessary for its successful realization. Nevertheless, there are still many practical challenges, i.e., interpreting digital data, deriving meaningful graphics, and subsequently scaling it, without information loss, to suit the perspective of an individual's visual field, ensuring precise low-latency tracking and monitoring, developing seamless optical displays, etc. All of these need to be addressed so that AR for next-generation ITS platforms can achieve its full potential.

## Cross-References

▶ Augmented Reality Entertainment: Taking Gaming Out of the Box
▶ Augmented Reality for Maintenance
▶ History of Virtual Reality
▶ Interaction with Mobile Augmented Reality Environments

**Acknowledgment** This publication received support from Science Foundation Ireland's CONNECT programme, and is co-funded under European Regional Development Fund's Grant Number 13/RC/2077.